\documentclass[12pt]{iopart}
\usepackage{graphicx}
\usepackage{iopams}
\usepackage{amssymb}
\usepackage{multirow}

\newcommand{\ped}[1]{\ensuremath{_{\rm #1}}}

\begin{document}
\title{Superconductive critical temperature of $Pb/Ag$ heterostructures}
\author{G.A.~Ummarino}
\ead{giovanni.ummarino@polito.it}
\address{Istituto di Ingegneria e Fisica dei Materiali,
Dipartimento di Scienza Applicata e Tecnologia, Politecnico di
Torino, Corso Duca degli Abruzzi 24, 10129 Torino, Italy; National Research Nuclear University MEPhI (Moscow Engineering Physics Institute),
Kashira Hwy 31, Moskva 115409, Russia}
\begin{abstract}
Recent experimental data (H. Nam et al., \textit{Phys. Rev. B} \textbf{100}, 094512 (2019)) of critical temperature and gaps
measured on superconductor/normal metal heterostructure ($Pb/Ag$), epitaxially grown, shown a interesting not usual behaviour.
The critical temperature decreases strongly but, despite the large differences in the lattice constants and electronic densities of states in the separate components, this heterostructure shows a spatially constant superconducting gap. In the paper it is demonstrated that the proximity Eliashberg equations, whit no free parameters, cannot explain the dependence of critical temperature from the rate $d_{Pb}/d_{Ag}$.
However it is sufficient to assume that the density of states at the Fermi level of silver is equal to that of lead in a layer adjacent to the separation interface, presumably of a thickness less than the coherence length of the lead, to perfectly explain the decrease in the critical temperature and the gap value in function of the rate $d_{Pb}/d_{Ag}$ always without free parameters.

\end{abstract}
%
%
\maketitle
\section{INTRODUCTION}
\label{intro}
The superconductive proximity effect is the modification of the superconducting properties of one metal when it is in contact with another \cite{degennes,Wolf}. The second metal may itself be superconductive, normal or characterized by special magnetic or other properties. The theory of the proximity effect usually presupposes a plain sample geometry \cite{McMillan}.
The large part of the experimental studies on proximity effect have been carried out using cryo-condensation thin films that are highly disordered with extremely short electron mean free paths.
Furthermore, their granular nature leads to a less well-defined interface, making it more difficult to understand the role of the interface in the understanding of this phenomenon.
Recent advancements in epitaxial growth of superconducting and nonsuperconducting metallic films with atomic precision have opened up new opportunities to investigate the
physics of the interface on the coupling efficiency in the proximity effect. More specifically, as the Fermi
surface of individual layers can be tuned with the thickness, the role of Fermi-surface matching and the individual electronic density of states can be directly estimated.
The author of the paper \cite{Nam} on $Pb/Ag$ heterostructure
find that their proximity system behaves as a single electronic system, rather than a two component
system and a uniform gap value persists through the $Pb/Ag$ interface. Such behavior is attributed to the nature of the interface.
In the paper I will try to explain the experimental data relative to $Pb/Ag$ heterostructure \cite{Nam} grown on Si(111) using molecular-beam epitaxy.

The paper is organized as follow. In Sec.~\ref{sec:model} I show the model I use for the computation of the superconductive critical temperature, i.e. the one band s-wave Eliashberg equations with proximity effect. In Sec.~\ref{sec:results} I discuss my results on $Pb/Ag$ heterostructure. Finally, conclusions are given in Sec.~\ref{sec:conclusions}.

\section{MODEL: PROXIMITY ELIASHBERG EQUATIONS}
\label{sec:model}
I calculated the critical temperature of the system and the two superconductive gaps in the $Pb$ and $Ag$ layers by solving the one band s-wave Eliashberg equations \cite{carbibastardo,ummarinorev} with proximity effect. In this case four coupled equations for the gaps $\Delta_{S(N)}(i\omega_{n})$ and renormalization functions $Z_{S(N)}(i\omega_{n})$ have to be solved (here $S$ and $N$ indicate ``superconductor" and ``normal" respectively and $\omega_{n}$ denotes the Matsubara frequencies). The Eliashberg equations with proximity effect on the imaginary-axis \cite{McMillan,Carbi1,Carbi2,Carbi3,Carbi4,kresin} are:
\begin{eqnarray}
&&\omega_{n}Z_{N}(i\omega_{n})=\omega_{n}+ \pi T\sum_{m}\Lambda^{Z}_{N}(i\omega_{n},i\omega_{m})N^{Z}_{N}(i\omega_{m})+\nonumber\\
&&+\Gamma\ped{N} N^{Z}_{S}(i\omega_{n})
\label{eq:EE1}
\end{eqnarray}
\begin{eqnarray}
&&Z_{N}(i\omega_{n})\Delta_{N}(i\omega_{n})=\pi
T\sum_{m}\big[\Lambda^{\Delta}_{N}(i\omega_{n},i\omega_{m})-\mu^{*}_{N}(\omega_{c})\big]\times\nonumber\\
&&\times\Theta(\omega_{c}-|\omega_{m}|)N^{\Delta}_{N}(i\omega_{m})
+\Gamma\ped{N} N^{\Delta}_{S}(i\omega_{n})\phantom{aaaaaa}
 \label{eq:EE2}
\end{eqnarray}
\begin{eqnarray}
&&\omega_{n}Z_{S}(i\omega_{n})=\omega_{n}+ \pi T\sum_{m}\Lambda^{Z}_{S}(i\omega_{n},i\omega_{m})N^{Z}_{S}(i\omega_{m})+\nonumber\\
&&\Gamma\ped{S} N^{Z}_{N}(i\omega_{n})
\label{eq:EE3}
\end{eqnarray}
\begin{eqnarray}
&&Z_{S}(i\omega_{n})\Delta_{S}(i\omega_{n})=\pi
T\sum_{m}\big[\Lambda^{\Delta}_{S}(i\omega_{n},i\omega_{m})-\mu^{*}_{S}(\omega_{c})\big]\times\nonumber\\
&&\times\Theta(\omega_{c}-|\omega_{m}|)N^{\Delta}_{S}(i\omega_{m})
+\Gamma\ped{S}N^{\Delta}_{N}(i\omega_{n})\phantom{aaaaaa}
 \label{eq:EE4}
\end{eqnarray}
where $\omega_{c}$ is a cutoff energy at least three times larger than the maximum phonon energy, $\mu^{*}_{S(N)}$ are the Coulomb pseudopotentials in the superconductive and normal layer respectively and $\Theta$ is the Heaviside function.
Moreover:
\begin{equation}
N^{\Delta}_{S(N)}(i\omega_{m})=\Delta_{S(N)}(i\omega_{m})/
{\sqrt{\omega^{2}_{m}+\Delta^{2}_{S(N)}(i\omega_{m})}}
\end{equation}
\begin{equation}
N^{Z}_{S(N)}(i\omega_{m})=\omega_{m}/{\sqrt{\omega^{2}_{m}+\Delta^{2}_{S(N)}(i\omega_{m})}}
\end{equation}
\begin{equation}
\Gamma_{S(N)}=\pi|t|^{2}Ad_{N(S)}N_{N(S)}(0)
\label{eq:EE6}
\end{equation}
with the relation $\frac{\Gamma_{S}}{\Gamma_{N}}=\frac{d_{N}N_{N}(0)}{d_{S}N_{S}(0)}$, where $A$ is the junction cross-sectional area, $|t|^{2}$ is the transmission matrix, $d_{S(N)}$ are the superconductive and normal layer thicknesses respectively, $N_{S(N)}(0)$ are the densities of states at the Fermi level for the superconductive and normal material.
Finally:
\begin{equation}
\Lambda_{S(N)}(i\omega_{n},i\omega_{m})=2
\int_{0}^{+\infty}d\Omega \Omega
\alpha^{2}_{S(N)}F(\Omega)/[(\omega_{n}-\omega_{m})^{2}+\Omega^{2}]
\end{equation}
where $\alpha^{2}_{S(N)}F(\Omega)$ are the electron-phonon spectral functions.
The electron-phonon coupling constants are defined as

\begin{equation}
\lambda_{S(N)}=2\int_{0}^{+\infty}d\Omega\frac{\alpha^{2}_{S(N)}F(\Omega)}{\Omega}\nonumber\\
\end{equation}

In order to solve this set of coupled equations, nine input parameters are needed: the two electron-phonon spectral functions $\alpha^{2}_{S(N)}F(\Omega)$, the two Coulomb pseudopotentials $\mu^{*}_{S(N)}$, the two values of the normal density of states at the Fermi level $N_{S(N)}(0)$, the thickness of the superconductive and normal layers $d_{S(N)}$ and the product between the junction cross-sectional area $A$ and transmission matrix $|t|^{2}$. The values of $d_{S(N)}$ , $A$  and  $|t|^{2}$ are experimental data. I assume the transmission matrix $|t|^{2}=1$ as I expect the interface between the surface and bulk layers to be nearly ideal
and $A=10^{-7}$ $m^{2}$. The final result is, of course, independent from $A$. The letter $S$ is for $Pb$ and the letter $N$ is for $Ag$ in the next, of course.
The electron-phonon spectral function of lead ($\lambda_{Pb}=1.55$) is present in literature \cite{carbibastardo} (it is shown in the insert of figure 1) as the value of the normal density of states at the Fermi level \cite{Ummarino_N0Pb} $N_{Pb}(0)=0.25611$ $eV^{-1}$ for unit cell while the value of the Coulomb pseudopotential is fixed for obtaining $T_{c}=7.22$ $K$ in a system without proximity effect: I find $\mu^{*}_{Pb}=0.133834$ by using a cutoff energy $\omega_{c} = 90$ meV and a maximum energy $\omega_{max} = 100$ meV.

The electron-phonon spectral function for the silver is unknown but it is possible to use the phonon density of states \cite{Xie_a2FAg} and normalized in order to have \cite{Allen_lambdaAg} $\lambda_{Ag}=0.10$.
This operation is correct because the electron-phonon coupling constant is week and it is reasonable to forecast that also its effect will be week (the silver spectral function is shown in the insert of figure 1).
For silver the value of the normal density of states at the Fermi level \cite{Butler_N0Ag} is $N_{Ag}(0)=0.13$ $eV^{-1}$ for unit cell while the value of the Coulomb pseudopotential \cite{Rapp_muAg} is $\mu^{*}_{Ag}=0.11$.
Now all input parameters of proximity Eliashberg equations are fixed so it is possible to calculate the critical temperature and superconductive gaps in function of rate $d_{Pb}/d_{Ag}$. This calculation has no free parameters.

\section{RESULTS AND DISCUSSION}
\label{sec:results}
If I solve the Eliashberg equations and I find the critical temperature in function of rate $d_{Pb}/d_{Ag}$ the agreement with experimental data is very bad (see black dashed line in figure 1).
So it is necessary to do some new hypothesis to solve this problem.
I assume that, in the case of almost ideal interfaces such as can be those of this experiment, \textit{in a normal metal layer of approximately equal thickness to that of the coherence length of the superconductor, the density of states at the Fermi level of the normal metal is replaced by the value present in the superconductive material.} The coherence length of lead \cite{coherencePb} is $960$ ${\AA}$ so the silver layer is always inside this distance.
In this way I solve the Eliashberg equation just by changing the value of $N_{Ag}(0)$: now $N_{Ag}(0)=N_{Pb}(0)$ but all other input parameters remain the same.  The result is the solid orange line in figure 1. Now the agreement is very good and the model has always no free parameters. In figure 2 are shown the calculated values of gaps in the superconductive and normal layers by solving the Eliashberg equation and by using Padè approximants.
It is possible to see that the gaps are almost coincident, exactly as it happens in the experiment and also the values are in perfect agreement with the experimental data.
\section{CONCLUSIONS}
\label{sec:conclusions}
The experimental data of $Pb/Ag$ heterostructures can be perfectly explained if it is assumed that, in the case of almost ideal interfaces such as can be those of this experiment, in a normal metal layer of approximately equal thickness to that of the coherence length of the superconductor, the density of the states at the Fermi level of the normal metal is replaced by the value present in the superconductive material. This mean that in the case of almost ideal interface if $N_{S}(0)>N_{N}(0)$ the critical temperature of system decreases more fast than in presence of a highly disordered film where the usual theory works. Of course if $N_{S}(0)<N_{N}(0)$ happen the opposite.
This assumption that allows to explain very well the experimental data could be a new stimulus for further theoretical investigations with models from first principles.
\ackn
G.A.U. acknowledges support from the MEPhI Academic Excellence Project (Contract No. 02.a03.21.0005).\\


%
\begin{figure}
\begin{center}
\includegraphics[keepaspectratio, width=\columnwidth]{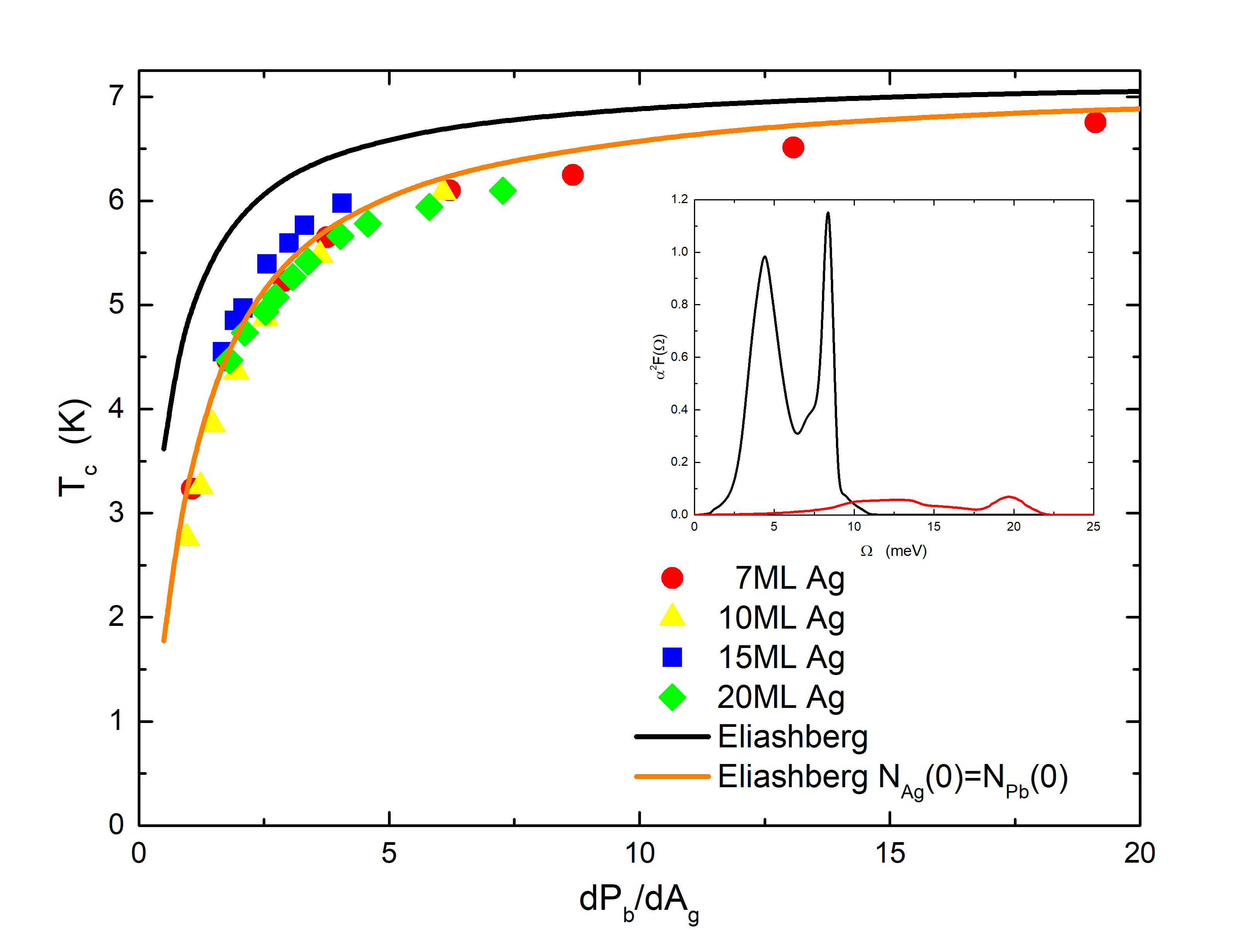}
\vspace{-5mm} \caption{(Color online)
 Calculated critical temperature by using different (black dashed line) and equal (orange solid line) values of densities of states at the Fermi level. The experimental data (full red circles, full yellow up triangles, full blue squares and full green rhombus) are from ref \cite{Nam}. In the inset the electron-phonon spectral functions of lead (black solid line) and silver (red solid line) are shown.
 }\label{Figure1}
\end{center}
\end{figure}
\newpage
\begin{figure}
\begin{center}
\includegraphics[keepaspectratio, width=\columnwidth]{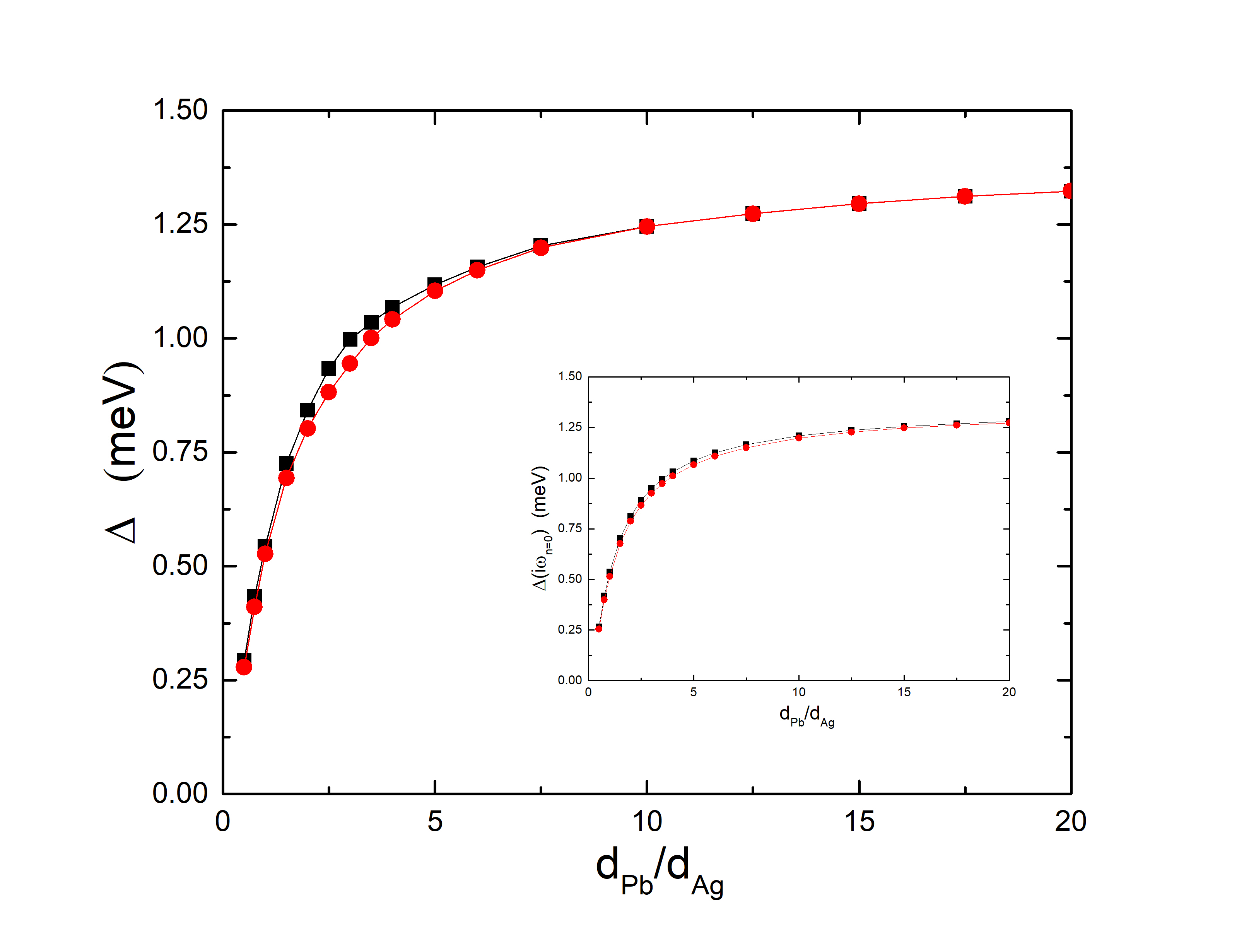}
\vspace{-5mm} \caption{(Color online)
 Calculated values of superconductive gaps in lead (full black squares) and in silver (full red circles) in function of rate $d_{Pb}/d_{Ag}$ by Padè approximants at $T=T_{c}/15$. In the inset it is shown the dependence
 always by the rate $d_{Pb}/d_{Ag}$ for $\Delta(i\omega_{n=0})$ for lead (full black squares) and silver (full red circles) obtained by solution of imaginary axis proximity Eliashberg equations. The lines are guides for the eye.
 }\label{Figure2}
\end{center}
\end{figure}

\end{document}